**Micaela Greene, MRC/CCI, Drexel University, mlg355@drexel.edu**
**Sam Grabus, MRC/CCI, Drexel University, smg383@drexel.edu**
**Jane Greenberg, MRC/CCI, Drexel University, jg3243@drexel.edu**

# DARSI: An Ontology for Facilitating the Development of Data Sharing and Use Agreements

**Abstract**
The advantages of data sharing across organizations and disciplines are indisputable; although, sensitive and restricted data cannot be easily shared due to policies and legal matters. The research presented in this paper takes a step toward systematizing the sharing of sensitive and restricted research data by developing an ontology to frame and guide DSUA (Data Sharing and Usage Agreement) development. The paper provides background context, describes the ontology creation process, and introduces the Data Sharing Agreements for Restricted and Sensitive Information (DARSI) ontology. DARSI contains four top level classes, 20 sub-classes, 33 sub-categories and 17 simple properties for categories applicable at various levels. The discussion provides further insight into the work accomplished, and the conclusion identifies next steps.

## 1. Introduction

The advantages of sharing data are indisputable. Combining research data from different projects for additional analyses can provide new perspectives, and increases the value of data. For example, medical research data combined with environmental data may give insight into important connections. A challenge is that not all data is open and easily shared due to policy and legal issues. This is a growing challenge as the pace of research and the volume of data generated increases daily, and much of this data is sensitive and restricted. To this end, in the United States the use and disclosure of sensitive and restricted information, including Personally Identifiable Information (PII), is regulated by laws that include the Health Insurance Portability and Accountability Act (HIPAA) (hhs.gov/hipaa) and other related laws and policies. Although such regulations are crucial for protecting individuals and enterprise knowledge, their application can be daunting for researchers lacking "legalese" expertise.

The most common solution is to develop a formal data sharing and usage agreement (DSUA) between the parties desiring to share data. DSUA's are highly structured documents; they specify the who, what, when, and how of data sharing, including which data attributes can be shared and how the data can be manipulated. Polanin and Terzian's (2018) research underscores the importance of a DSUA for protecting PII data, conducting responsible and ethical research, and increasing researchers' willingness to share data. A challenge here is that DSUA's are complex legal agreements; they are costly to create in terms of time and legal fees, and this predicament ultimately hinders timely sharing of useful data.

Data sharing challenges, while frustrating, also point to the need to develop more systematized, even automatic processes for producing DSUA's. This need underlies the overall goal of the research reported on in this paper. Specifically, we report on work to develop an ontology as an overall knowledge structure that can frame and guide DSUA development. The sections to follow provide background context for our research conducted as part of the NSF Northeast Big Datahub Spoke project, "A Licensing Model and Ecosystem for Data Sharing." The goals of our work and the Data Sharing Agreements for Restricted and Sensitive Information (DARSI) ontology are presented, followed by a brief discussion and conclusion.

## 2. Background Research

The FAIR principles supporting data sharing and the role of data sharing systems and infrastructure inform our approach in addressing the challenges reviewed in the introduction.

*FAIR Principles*

The FAIR Data Principles (Wilkinson et al. 2016) hold that research data should be Findable, Accessible, Interoperable, and Reusable. The principles are interconnected with the open data movement, which was motivated by a number of converging factors (e.g. launch of the World Wide Web, open source software development, open access, federal data sharing policies, etc.) (Tenopir et al. 2011). Open data allows for research verification; and, perhaps more significantly, open data increases data access, use, reuse, and repurposing to advance knowledge and innovation. The FAIR principles promote these goals, but also provide context for data that may be sensitive and restricted, often referred to as "closed data." Closed data may contain private, sensitive, or even competitive information about a person, organization, issue, activity, or other factor, and it is generally protected by an authorized policy or law.

Data that is closed, however, can still be guided by FAIR principles within the allowable context. For example, closed data should still be "Findable" by those who need to find it, and it should be "Accessible," "Interoperable," and "Reusable" in the closed use environment. The FAIR principles facilitate good, sustainable data management practices, ensuring current and future use. Too often, data generated for an immediate need cannot be found, accessed, or used at a later date due to poor data management practices. Supporting FAIR can help address this problem. The FAIR principles also serve technical system processes and infrastructure, and are reviewed below.

*Data Sharing Systems and Infrastructure*

Open access and open data have vastly changed the information sharing landscape over the last two decades, due to open source repository software development, the growth in open data repositories, and the adoption of shared metadata standards. These systems provide an infrastructure for supporting FAIR principles, although they do not fully support FAIR as applied to closed data. This is because the necessary protections and regulations are not fully in place as outlined in the DSUA. Indeed, there are key developments, such REDCap (projectredcap.org/software/) used for sharing data on research teams in hospitals, and Dataverse (dataverse.org/), which has implemented a datatag system that indicates security features and requirements, and access requirements for handling data (Sweeney, et al. 2015). These exemplary systems are important developments, although they are not designed to generate a legally binding DSUA, which is required particularly when sharing data involves different institutions.

The use of ontology expert systems present an example of how we can apply the use of ontologies to DSUA development (e.g. Han and Park 2009). Ontologies have also been used to address privacy concerns in medical research data (Li and Samavi 2018), smart technology (Toumia, Szoniecky, and Saleh 2020), and other matters, further informing our work. Additionally, we can turn to Vivli (vivli.org/), a data sharing platform supporting global access to clinical research. Vivli's workflow guides researchers in completing a data contribution form, a data contribution agreement, and confirming data anonymization. The Vivli template includes standard language for the agreements, and researchers sign this document, although it is limited to data sharing specific to clinical trials. Drawing on all these developments and the need to systematize DSUA creation more broadly, we have pursued the development of an ontology to guide this work.

## 3. Goals

The overall goal of our research is to develop an ontology that can simplify the complex process of generating DSUA's, particularly when sensitive and restricted data is involved. The overall research goal was guided by the following three objectives:

1. Identify common data attributes found across DSUA's involving sensitive and restricted data.
2. Systemize the workflow for applying these data attributes.
3. Arrange the common data attributes (from objective 1) in an ontological framework reflecting a systematized workflow (mapped out in objective 2).

The ontology developed is intended to accelerate and boost collaboration between entities while enforcing standardized data sharing contract terms. The methods, data sample, and steps for this work reported on in the next section.

## 4. Methods and Approach

General ontology engineering methodology was used to guide our work. The approach included sample collection development, a set of analyses, ontology development, and ontology encoding.

1. *Sample collection*: A sample of 80 documents containing both completed DSUA's and DSUA templates was gathered. The sample includes the 30 agreements used in Grabus and Greenberg (2017). The completed DSUA's and templates emphasize academic and industry partnerships, although education, government, and non-profits are also represented.
2. *Analyses*: Automatic text extraction and human examination were combined to identify components that are generally included in a comprehensive agreement. A contextual analysis followed. This step involved consulting additional data guidelines and resources, including the *National Neighborhood Indicators Partnership* (NNIP 2018) and *Contracts for Data Collaboration* (C4DC 2019), to assess the core elements for the developing ontology..
3. *Ontology development*: The language components (elements) were parsed and arranged into higher-level general categories, and by mid and lower-level attributes. The high-level general categories and lower-level attributes formed a hierarchical structure. This step was conducted in conjunction with step two, as a back-and-forth activity for verification.
4. *Ontology encoding*. The attribute set was encoded using the Protégé ontology tool. Protégé supported establishing classes, subclasses, and attributes in the Web Ontology Language (OWL). These classes reflect significant elements of data sharing agreements.

The full collection of categories and attributes that underlies the DARSI Ontology is presented below in the results section.

## 5. Results

The steps outlined in section four result in the development of an ontological framework of classes, subclasses, and properties of the attributes, and describe the relationships among the concepts and elements. DARSI includes four top classes⸺ Contract Terms & Conditions, Responsible Parties, Project Description, and Data Handling⸺and 20 subclasses (Figure 1). DARSI also includes 33 sub-categories of the subclasses and 17 simple properties for categories at various levels.

When considering the lower-level terms, we classified them either as an attribute or as a property. Concepts that represent complex ideas and require detailed specifications are classified as attributes. Examples of attributes include methods put in place for ensuring compliance with regulations, and conditions under which the data can be manipulated; both are subclass attributes of the Data Handling top class. Properties can be specified in just a few words. For example, properties belonging to the *data description* subclass of the Project Description top class include simple terms like collection dates, data format, and data owner(s).

Figure 1: DARSI Ontology Top Classes and Subclasses

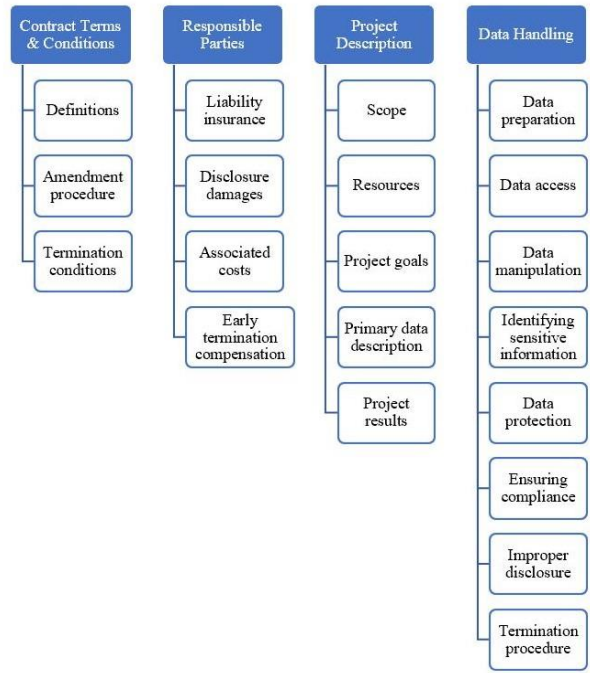

The overall ontology was examined in a visualization application called VOWL, which supports OWL and allows us to explore the relationships. The following figures provide a visualization of these relationships, focusing on a set of classes and subclasses within the Project Description (Figure 2) and Data Handling (Figure 3) classes.

Figure 2: Detail of DARSI Ontology Project Description Top and Subclasses
Figure 3: Detail of DARSI Ontology Data Handling Top and Subclasses

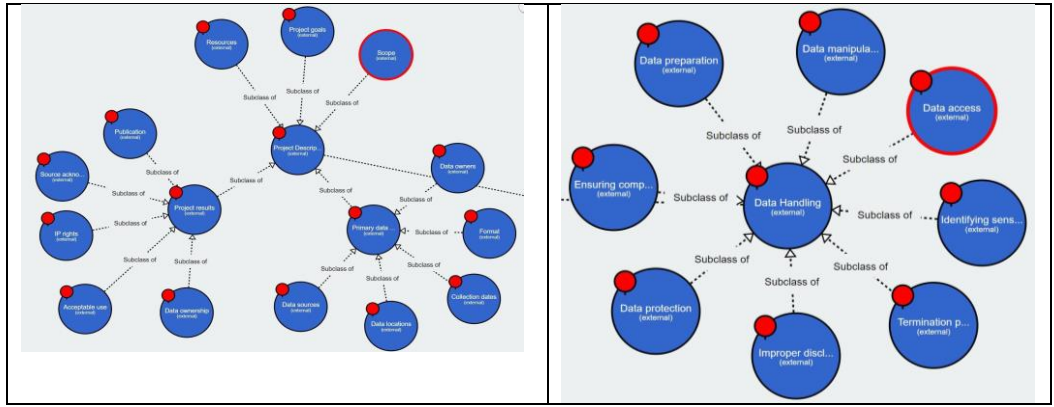

## 6. Discussion

The development of DARSI allowed us to move from a strict hierarchy to an ontological model, and express more complex relationships that are indicative of sharing sensitive and restricted data. Specifically, we pulled apart the initial hierarchical structure and examined each term and its relationships within the wider scope of our ontology goals. The ability to include ontology relationships enriches understanding of the data sharing domain by encoding and representing connections among the elements, both within and across categories.

The data use agreement ontologies created by other researchers—including DSAP, focused on medical research (Li and Samavi 2018), and ColPri (Toumia, Szoniecky, and Saleh 2020), oriented toward privacy concerns of individuals—informed and inspired our work on DARSI. Our sample DSUA's included many academic and industry partnerships; as a result we found attributes that were common to the other ontologies and also identified additional ones. With a particular interest in the technical aspects of how data of any type is handled and processed, we aimed to develop the subclasses and properties of the Project Description and Data Handling top classes as exhaustively as possible.

## 7. Conclusion

At present, most DSUA's are created from the ground up. Each agreement has internal consistency; however, unique agreements hinder the development of a broader ecology for data sharing and use on a larger scale, across disciplines. A standard and widely available licensing model could aid in systematizing, streaming, and even automating the process of generating a DSUA to facilitate data sharing in a timely and secure manner. This need motivated the work presented in this paper, and has also informed future work. Specifically, we will pursue two key steps.

Develop DARSI 2.0. The DARSI ontology presented here is an initial rendering. The team recognizes the need to consult with experts in legal contract language and further review our framework. To this end, an attorney has joined our team, and will be reviewing the language over the coming months.

Review for cross-disciplinary and transdisciplinary language. As noted, there are similar initiatives exclusive to sharing sensitive and restricted data in specific domains such as medical research. DARSI has been generated as an ontology to support sharing across disciplines and environments. We intend to seek review from cross-disciplinary and transdisciplinary researchers who have engaged in sharing sensitive and restricted data. We will investigate the impacts of keeping the ontology more general, or potentially generating more specialized versions of the ontology to meet certain domain criteria.

Overall, the DARSI ontology provides a useful knowledge framework for helping to address a growing problem with delays in sharing sensitive and restricted research data. The ontological engineering approach used, and pulling apart the initial hierarchical rendering, can be helpful in developing ontologies intended to help systemize or complex workflows. Finally, the results presented can inform future system application development.